\documentclass[12pt]{article}

\usepackage{latexsym}

\begin{document}


\vskip 1truecm

\title{Tensor mass and particle number peak at the same location in the
scalar-tensor gravity boson star models - an analytical proof}

\author{Yazadjiev S.${}^{*}$\\
\\
{\footnotesize ${}^{*}$ Department of Theoretical Physics,
Faculty of Physics,}\\
{\footnotesize Sofia University,}\\
{\footnotesize 5 James Bourchier Boulevard, Sofia~1164, }\\
{\footnotesize Bulgaria }\\
{\footnotesize  E-mail: yazad@phys.uni-sofia.bg}
}

\maketitle

\begin{abstract}
Recently in boson star models in framework of Brans-Dicke theory,
three possible  definitions  of mass  have been identified, all
identical in general relativity, but different in scalar-tensor theories
of gravity.It has been conjectured  that it's the tensor mass which
peaks, as a function of the central density, at the same location
where the particle number takes its maximum.This is  a very important
property  which is crucial for stability analysis via catastrophe theory.
This conjecture has received some numerical support.
Here we give an analytical proof of the conjecture in framework of the
generalized scalar-tensor theory of gravity, confirming in this way
the numerical calculations.
\end{abstract}


\sloppy
\renewcommand{\baselinestretch}{1.3} %
\newcommand{\sla}[1]{{\hspace{1pt}/\!\!\!\hspace{-.5pt}#1\,\,\,}\!\!}
\newcommand{\db}{\,\,{\bar {}\!\!d}\!\,\hspace{0.5pt}}
\newcommand{\partb}{\,\,{\bar {}\!\!\!\partial}\!\,\hspace{0.5pt}}
\newcommand{\dsla}{\partb}
\newcommand{\eql}{e _{q \leftarrow x}}
\newcommand{\eqr}{e _{q \rightarrow x}}
\newcommand{\ite}{\int^{t}_{t_1}}
\newcommand{\itz}{\int^{t_2}_{t_1}}
\newcommand{\itd}{\int^{t_2}_{t}}
\newcommand{\lfrac}[2]{{#1}/{#2}}
\newcommand{\sfrac}[2]{{ \,\,\hbox{${\frac {#1} {#2}}$}}}
\newcommand{\dV}{d^4V\!\!ol}
\newcommand{\ben}{\begin{eqnarray}}
\newcommand{\een}{\end{eqnarray}}
\newcommand{\la}{\label}

\vskip 1cm
Boson stars were first discussed by Kaup \cite{K} and then by Ruffini and
Bonazzola \cite{RB}.
Boson stars in scalar-tensor theories of gravity have been investigated
extensively by many researchers.The first model of a boson star
in pure Brans-Dicke theory has been studied by Gunderson and Jensen \cite{GJ}.
Their work was generalized by Torres \cite{T} who was studied boson stars in
scalar-tensor theories with non-constant $\omega_{BD}(\Phi)$.More recently
boson stars have investigated in the papers by Torres et al.
\cite{TLS},\cite{TSL} and in the paper by Comer and Shinkai \cite{CS}.
In \cite{TLS} boson stars have been studied in connection with so-called
gravitational memory \cite{Barrow}, while their stability through cosmic
history has examined in \cite{CS} and \cite{TSL}.Finally, the dynamical
evolution of boson stars has been investigated in the paper by
Balakrishna and Shinkai \cite{BS}.
For more details we refer the reader to the most recent review on boson
stars \cite{MS}.

Here we consider complex scalar field boson stars in the most general scalar
tensor theory of gravity with an action in Jordan frame
\ben
S= -{1 \over 16\pi G_{*}}\int \sqrt{-\tilde g}\left(F(\Phi)\tilde R -
H(\Phi){\tilde g}^{\mu\nu}
\partial_{\mu}\Phi\partial_{\nu}\Phi + \tilde U(\phi)\right)d^4x   + \\
\nonumber
+ \int \sqrt{-\tilde g}\left({1\over 2}{\tilde g}^{\mu\nu}
\partial_{\mu}\Psi^{+}\partial_{\nu}\Psi - W(\Psi^{+}\Psi)\right)d^4x
\la{JA}
\een
where $\tilde R$ is Ricci scalar curvature with respect to the space-time
metric $\tilde g_{\mu\nu}$, $G_{*}$ is the bare Newtonian constant and
$\Phi$ is the gravitational Brans-Dicke scalar with potential term
$\tilde U(\phi).$  $\Psi$ is massive  self-interacting complex scalar field
with
$$W(\Psi^{+}\Psi) = {1\over 2}m^2\Psi^{+}\Psi  +
{1\over 4}\lambda_{*} (\Psi^{+}\Psi)^2.$$

Hereafter we will consider only static and spherically symmetric boson
stars.

The explicit form of the action shows that we have $U(1)$-invariant action
under global gauge transformation $\Psi\to e^{ia}\Psi $, $a$ being
a constant.This global $U(1)$-symmetry gives rise to the following conserved
current
\ben
J^{\mu}={i\over 2}\sqrt{-\tilde g}
{\tilde g}^{\mu\nu}\left(\Psi\partial_{\nu}\Psi^{+} -
\Psi^{+}\partial_{\nu}\Psi\right)
\la{JC}
\een
The conserved current leads to a conserved charge - total particle number
making up the star
\ben
N = \int J^{0}d^3x
\een
Binding energy is then defined by
\ben
E_{B}=M_{Star} - m N
\een
where $m$ is the particle mass.

Now a problem arises: How to define the mass which appears in the
expression for the binding energy?

In contrary to general relativity the definition of mass in scalar-tensor
theories of gravity is quite subtle.This problem has been recently examined
numerically in the work by Whinnett \cite{Whinnett}.He has considered three
possible mass definitions in Jordan frame, namely Schwarzschild mass $M_{S}$
(i.e. ADM mass in Jordan frame),
Keplerian  mass $M_{Kepler}$ and the tensor one $M_{T}$.As it has been shown
numerically in \cite{Whinnett} (see also \cite{BFY}) these three masses
differ significantly from each other in the case $\omega_{BD} =-1$.Keplerian
mass leads to positive binding energy which means that every boson star
solution is in general potentially unstable.Contrary, Schwarzschild mass
leads to negative binding energy  suggesting that every solution is
potentially stable even for large  central densities $\rho=\mid\Psi(0)\mid^2$.

It should be noted that for large constant $\omega_{BD}$
(say $\omega_{BD} > 500$)
the difference between the three masses is negligible.However, for arbitrary
\footnote{We mean arbitrary $\omega_{BD}(\phi)$ for which the theory passes
through all known gravitational experiments.}
$\omega_{BD}(\phi)$  it's possible
that the three masses may differ from each other significantly.This may
occur in the early universe  when the cosmological value
$\Phi_{\infty}$ is sufficiently smaller than $1$ \cite{TSL}.
Moreover, our numerical calculations show that for some physically relevant
functions $\omega_{BD}(\phi)$  we may have
$M_{T} -  M_{S}\approx (0.15-0.20)M_{T}.$
On the other hand the scalar tensor theories of gravity with
$\omega_{BD}=-1$ better describe the early universe(see  \cite{TW} and
references therein).So, the case when $\omega_{BD}$  is not large,
is also physically relevant.Therefore, when we study the boson stars in
the early universe the mass choice is crucial.

It's the tensor mass which leads to physically acceptable picture.In
\cite{Whinnett}  it's shown numerically that the tensor mass peaks
at the same point as particle numbers  - a very important property in
general relativity \cite{HTWW}.This property is also crucial for the
application of catastrophe theory to analyze the stability of the
boson stars \cite{KMS1},\cite{KMS2}.

In \cite{Whinnett} and \cite{TSL} the problem for analytical proof of
the fact that it's the tensor mass which peaks at the same location
as the particle number has been stated. The main purpose of this paper
is to fill this gap.

In our opinion it's more convenient to work in Einstein frame given by
\ben
g_{\mu\nu}=F(\Phi)\tilde g_{\mu\nu}
\een
In Einstein frame the action (\ref{JA}) takes the form
\ben
\la{EA}
S=-{1 \over 16\pi G_{*}}\int \sqrt{-g}\left(R -
2g^{\mu\nu}
\partial_{\mu}\phi\partial_{\nu}\phi  + U(\phi)\right)d^4x   +  \\ \nonumber
+ \int \sqrt{-g}\left({1\over 2}A^2(\phi){g}^{\mu\nu}
\partial_{\mu}\Psi^{+}\partial_{\nu}\Psi -A^4(\phi)
W(\Psi^{+}\Psi)\right)d^4x
\een
where $R$ is the Ricci scalar curvature with respect to the metric
$g_{\mu\nu}$, $U(\phi)=A^4(\phi)\tilde U(\Phi(\phi))$
and $A^2(\phi)= F^{-1}(\Phi(\phi))$ as $\phi$ is given by
\ben
\phi = \int d\xi \sqrt{ {3\over 4} \left({d\ln{(F(\xi))}\over d\xi}\right)^2
+ {1\over 2}{H(\xi)\over F(\xi)}}
\la{Ff}
\een
The action (\ref{EA}) leads to the following fields equations
\ben
G_{\mu}^{\nu}=\kappa_{*}T_{\mu}^{\nu} +
2\partial_{\mu}\phi\partial^{\nu}\phi  -
\partial^{\sigma}\phi\partial_{\sigma}\phi\delta_{\mu}^{\nu} +
{1\over 2}U(\phi)\delta_{\mu}^{\nu} \\ \nonumber
\Box\phi + {1\over 4}U^{\prime}(\phi) = -{\kappa_{*}\over 2}\alpha(\phi)T
\\ \nonumber
\Box\Psi + 2\alpha(\phi)\partial^{\sigma}\phi\partial_{\sigma}\Psi =
-2A^2(\phi){\partial W(\Psi^{+}\Psi)\over \partial \Psi^{+}}
\\ \nonumber
\Box\Psi^{+} + 2\alpha(\phi)\partial^{\sigma}\phi\partial_{\sigma}\Psi^{+}
= -2A^2(\phi){\partial W(\Psi^{+}\Psi)\over \partial \Psi^{+}}
\een
where $\Box$ is d'Alambert operator in terms of the metric $g_{\mu\nu}$,
$\kappa_{*}=8\pi G_{*}$,$\alpha(\phi)={d\over d\phi}\ln(A(\phi))$
and $T$ is the trace of the Einstein frame
energy-momentum tensor of the complex scalar field  given by
\ben
T_{\mu}^{\nu} = {1\over 2}A^2(\phi)
\left(\partial_{\mu}\Psi^{+}\partial^{\nu}\Psi +
\partial_{\mu}\Psi\partial^{\nu}\Psi^{+}\right) -  \\ \nonumber
- {1\over 2}A^2(\phi)\left(\partial_{\sigma}\Psi^{+}\partial^{\sigma}\Psi
- 2A^2(\phi)W(\Psi^{+}\Psi)\right)\delta_{\mu}^{\nu}
\een

The conserved $U(1)$-current in Einstein frame is
\ben
J^{\mu}={i\over 2}A^2(\phi)\sqrt{-g}g^{\mu\nu}
\left(\Psi\partial_{\nu}\Psi^{+} -  \Psi^{+}\partial_{\nu}\Psi\right)
\la{ECC}
\een

As we have already mentioned we consider static and spherically symmetric
boson stars i.e. space-time with a line element in Einstein frame
\ben
ds^2 = e^{\nu}dt^2 - e^{\lambda}dr^2 - r^2\left(d\theta^2 +
\sin^2(\theta)d\varphi^2\right)
\een
and complex scalar field in the form
\ben
\Psi = \sigma(r)e^{i\omega t}
\een
where $\omega$ is real positive number and  $\sigma(r)$  is a real function.
It should be noted that the asymptotic behaviour of the field
$\phi$  set the following tight constraint
\ben
     0 < \omega < m
\een
In this case the field equation system reduces to the following system
of ordinary differential equations
\ben
\la{ODES}
\lambda^{\prime}={1 - e^{\lambda}\over r} + \kappa_{*}e^{\lambda}rT_{0}^{0}
+ r{\phi^{\prime}{}}^2  + {1\over 2}U(\phi)e^{\lambda}r \\ \nonumber
\nu^{\prime}= {e^{\lambda}-1\over r} - \kappa_{*}e^{\lambda}rT_{1}^{1}
+ r{\phi^{\prime}{}}^2  - {1\over 2}U(\phi)e^{\lambda}r  \\
\nonumber
\phi^{\prime\prime} = -\left({\nu^{\prime}-\lambda^{\prime}\over 2}
+{2\over r}\right)\phi^{\prime}  + {1\over 4}U^{\prime}(\phi)e^{\lambda}
+ {\kappa_{*}\over 2}\alpha(\phi)Te^{\lambda} \\ \nonumber
\sigma^{\prime\prime} = -\left({\nu^{\prime}-\lambda^{\prime}\over 2}
+{2\over r}\right)\sigma^{\prime}  - \omega^2 e^{\nu - \lambda}\sigma -
2\alpha(\phi)\phi^{\prime}\sigma^{\prime}  +
2A^2(\phi)e^{\lambda}W^{\prime}(\sigma^2)\sigma
\een
Here the components $T_{0}^{0}$ and $T_{1}^{1}$  are given correspondingly
by
\ben
T_{0}^{0}={1\over 2}\omega^2A^2(\phi)e^{-\nu}\sigma^2  +
{1\over 2}A^2(\phi)e^{-\lambda}{\sigma^{\prime}}^2  + A^4(\phi)W(\sigma^2)
\een

\ben
T_{1}^{1}=-{1\over 2}\omega^2A^2(\phi)e^{-\nu}\sigma^2 -
{1\over 2}A^2(\phi)e^{-\lambda}{\sigma^{\prime}}^2  + A^4(\phi)W(\sigma^2)
\een
The system (\ref{ODES}) has to be solved at the following boundary
conditions.We demand asymptotic flatness which means that $\nu(\infty)=0$.
On the other hand  nonsingularity at the origin requires $\lambda(0)=0$.
Concerning $\phi$,nonsingularity at the origin implies $\phi^{\prime}(0)=0$
while at infinity $\phi$ has to match the cosmological value
$\phi(\infty)=\phi_{\infty}$.Nonsingularity of $\sigma$ at the  origin
implies $\sigma^{\prime}(0)=0$.We require finite mass an therefore we
put $\sigma(\infty)=0$.In addition we have to give the central
value $\sigma(0)$.

It's well known that the tensor mass is just the ADM mass in Einstein frame
(we note that in Einstein frame all mass definitions coincide).Therefore
we can write directly the explicit expression for the tensor mass
using the first equation of (\ref{ODES}), namely
\ben
M_{T}={1\over 2G_{*}}\int_{0}^{\infty}dr r^2 \left(\kappa_{*}T_{0}^{0}
+ e^{-\lambda}{\phi^{\prime}}^2 + {1\over 2}U(\phi)\right)=
{1\over 2G_{*}}\int_{0}^{\infty}dr r^2 {\cal D}(r)
\la{TM}
\een
as ${\cal D}(r)$  is define by the expression itself.Respectively, the
particle number is given by
\ben
N=4\pi\int_{0}^{\infty}dr r^2e^{{\lambda}\over 2}\left(\omega A^2(\phi)
\sigma^2 e^{-{\nu\over 2}}\right)
\la{PN}
\een
When we are interested only in zeronodes solutions the boson star states are
parameterized by the $\sigma$ field central value $\sigma(0)$ or
equivalently by $\rho = \sigma^2(0).$ Let's consider
two infinitesimally nearby configurations parameterized by
$\sigma(0)$ and $\sigma(0) + \delta\sigma(0)$.The corresponding variations
of the tensor mass and particle number are
\ben
\la{DMT1}
\delta M_{T} = {1\over 2G_{*}}\int_{0}^{\infty}dr r^2 \delta {\cal D}
\een

\ben
\la{DN1}
\delta N = 4\pi\int_{0}^{\infty}dr r^2 \delta
\left(e^{{\lambda \over 2}}\omega A^2(\phi)\sigma^2 e^{-{\nu\over 2}}\right)
=  \\ \nonumber
4\pi\int_{0}^{\infty}dr r^2 e^{{\lambda \over 2}}\delta
\left(\omega A^2(\phi)\sigma^2 e^{-{\nu\over 2}}\right)   +
4\pi\int_{0}^{\infty}dr r^2
\left(\omega A^2(\phi)\sigma^2 e^{-{\nu\over 2}}\right)
\delta e^{{\lambda \over 2}}
\een
It's not difficult to obtain
\ben
\delta e^{{\lambda \over 2}}={1\over 2r}e^{{3\over 2}\lambda}
\int_{0}^{r}dr r^2 \delta {\cal D}
\la{DL}
\een
Now substituting (\ref{DL}) in (\ref{DN1}) and after some algebra
we have
\ben
\la{DN2}
{\delta N \over 4\pi} =\int_{0}^{\infty}dr r^2 e^{{\lambda \over 2}}\delta
\left(\omega A^2(\phi)\sigma^2 e^{-{\nu\over 2}}\right)  + \\ \nonumber
+ {1\over 2}\int_{0}^{\infty}dr r^2 \delta {\cal D}
\int_{r}^{\infty}d\xi \xi
\left(\omega A^2(\phi)\sigma^2 e^{-{\nu\over 2}}\right)e^{{3\over 2}\lambda}
\een
Taking into account the explicit form (\ref{TM}) of $\cal D$   one  obtains
\ben
\la{FS1}
\delta\left(\omega A^2(\phi)\sigma^2 e^{-{\nu\over 2}}\right)=
{e^{{\nu\over 2}} \over \omega}
\left(\delta\left({{\cal D}\over \kappa_{*}}\right) +
{1\over 2}\omega^2 e^{-\nu}\delta\left(A^2(\phi)\sigma^2\right)\right) \\
\nonumber
- {e^{{\nu\over 2}} \over \omega}
\delta\left({1\over 2}A^2(\phi)e^{-\lambda}{\sigma^{\prime}}^2
+ A^4(\phi)W(\sigma^2) + {1\over \kappa_{*}}e^{-\lambda}{\phi^{\prime}}^2
+ {1\over 2\kappa_{*}}U(\phi)\right)
\een
In more detailed form the expression (\ref{FS1}) is written as follows
\ben
\la{FS2}
\delta\left(\omega A^2(\phi)\sigma^2 e^{-{\nu\over 2}}\right)=
{e^{{\nu\over 2}} \over \omega}\delta\left({{\cal D}\over \kappa_{*}}\right)
- {e^{{\nu\over 2}} \over \omega}\left(L_{\phi}\delta\phi +
{2\over \kappa_{*}}e^{-\lambda}\phi^{\prime}\delta\phi^{\prime}\right) \\
\nonumber
- {e^{{\nu\over 2}} \over \omega}\left(L_{\sigma}\delta\sigma +
A^2(\phi)e^{-\lambda}\sigma^{\prime}\delta\sigma^{\prime}\right)
- {e^{{\nu\over 2}} \over \omega}
\left({1\over 2}A^2(\phi){\sigma^{\prime}}^2  +
{1\over \kappa_{*}}{\phi^{\prime}}^2\right)\delta e^{-\lambda}
\een
where $L_{\phi}$ and $L_{\sigma}$  are given by
\ben
L_{\phi} =\alpha(\phi)\left(-\omega^2 e^{-\nu}A^2(\phi)\sigma^2 +
A^2(\phi)e^{-\lambda}{\sigma^{\prime}}^2  + 4A^4(\phi)W(\sigma^2)\right) +
{1\over 2\kappa_{*}}U^{\prime}(\phi)
\een

\ben
L_{\sigma} = -\omega^2 e^{-\nu}A^2(\phi)\sigma  +
2A^4(\phi)W^{\prime}(\sigma^2)\sigma
\een
Putting (\ref{FS2}) in the first integral of (\ref{DN2}), performing
integration by parts and taking into account the third and the fourth
equation of the system (\ref{ODES}) one arrives at
\ben
\la{FS3}
\int_{0}^{\infty}dr r^2 e^{{\lambda \over 2}}\delta
\left(\omega A^2(\phi)\sigma^2 e^{-{\nu\over 2}}\right) =
\int_{0}^{\infty}dr r^2 {e^{{1\over 2}(\nu + \lambda)}\over \omega}
\delta\left({{\cal D}\over \kappa_{*}}\right) - \\ \nonumber
-\int_{0}^{\infty}dr r^2 {e^{{1\over 2}(\nu + \lambda)}\over \omega}
\left({1\over 2}A^2(\phi){\sigma^{\prime}}^2  +
{1\over \kappa_{*}}{\phi^{\prime}}^2\right)\delta e^{-\lambda}
\een
Substituting now (\ref{FS3}) in  (\ref{DN2}) as one uses that
$$\delta e^{-\lambda}= -{1\over r}\int_{0}^{r}dr r^2 \delta {\cal D}$$
we have
\ben
{\delta N \over 4\pi} =\int_{0}^{\infty}dr r^2
\Lambda(r) \,\, \delta \left({{\cal D}\over \kappa_{*}}\right)
\een
where $\Lambda(r)$ is the following expression
\ben
\Lambda(r) =
\kappa_{*}\int_{r}^{\infty} d\xi \xi
\left({1\over \omega}e^{{1\over 2}(\nu + \lambda)}
\left({1\over 2}A^2(\phi){\sigma^{\prime}}^2 +
{1\over \kappa_{*}}{\phi^{\prime}}^2 \right)  +
{1\over 2}\omega A^2(\phi)\sigma^2
e^{-{\nu\over 2}}e^{{3\over 2}\lambda}\right) +
\\ \nonumber
+{1\over \omega}e^{{1\over 2}(\nu + \lambda)}
\een
Using the first two equations of (\ref{ODES}), it's not difficult one to
show that $\Lambda$ is actually a constant which turns out to be
$\Lambda={1\over \omega}$.Therefore, we obtain finally
\ben
{\delta N \over 4\pi} ={1\over \omega}\int_{0}^{\infty}dr r^2
\delta {{\cal D}\over \kappa_{*}}
\een
Comparing this expression with (\ref{DMT1}) we conclude that the following
important relation holds
\ben
\delta M_{T} =\omega \delta N
\la{MTN}
\een
The relation (\ref{MTN}) may be rewritten  in the form
\ben
{\delta M_{T}\over \delta \rho} = \omega {\delta N\over \delta \rho}
\la{DMTN}
\een
Taking into account that $0< \omega< m$,
it follows from  (\ref{DMTN}) that if $\rho_{crit}$  is a critical point
for $M_{T}$  (i.e. ${\delta M_{T}\over \delta \rho} = 0$), then
$\rho_{crit}$ is also critical point for  $N.$

Let $M_{T}$ has a maximum  at $\rho = \rho_{crit}$
(${{\delta}^2 M_{T}\over \delta {\rho}^2} < 0$).Then we obtain
\ben
\left({\delta}^2 N\over\delta{\rho}^2\right)_{crit} ={1 \over \omega_{crit}}
\left({\delta}^2 M_{T}\over \delta {\rho}^2\right)_{crit} < 0
\een
which shows that $N$ has a maximum at $\rho=\rho_{crit}$, too.

Therefore the tensor mass $M_{T}$ and the particle number $N$ peak,
as functions of the central density, at the same location .

That the tensor mass and particle number peak at the same location results
in a cusp in the bifurcation diagram $M_{T}$ versus $N.$
In infinitesimal small neighborhood of a cusp we may take expansions in
powers of $(\rho - \rho_{crit})$
\ben
M_{T} = {M_{T}}_{crit}  +
{1\over 2}\left({\delta}^2 M_{T}\over \delta {\rho}^2\right)_{crit}
(\rho - \rho_{crit})^2  +
{1\over 6}\left({\delta}^3 M_{T}\over \delta {\rho}^3\right)_{crit}
(\rho - \rho_{crit})^3   + O(4) \\
\nonumber
N = {N}_{crit}  +
{1\over 2}\left({\delta}^2 N\over \delta {\rho}^2\right)_{crit}
(\rho - \rho_{crit})^2  +
{1\over 6}\left({\delta}^3 N\over \delta {\rho}^3\right)_{crit}
(\rho - \rho_{crit})^3   + O(4)
\een
Now using (\ref{DMTN}), the coefficients
$\left({\delta}^2 M_{T}\over \delta {\rho}^2\right)_{crit}$  and
$\left({\delta}^3 M_{T}\over \delta {\rho}^3\right)_{crit}$
may be expressed by
$N^{\prime\prime}=\left({\delta}^2 N\over \delta {\rho}^2\right)_{crit}$  and
$N^{\prime\prime\prime}=
\left({\delta}^3 N\over \delta {\rho}^3\right)_{crit}$, and the result is
$$\left({\delta}^2 M_{T}\over \delta {\rho}^2\right)_{crit} =
\omega_{crit}N^{\prime\prime}$$
$$\left({\delta}^3 M_{T}\over \delta {\rho}^3\right)_{crit}
=\omega_{crit}N^{\prime\prime\prime}  +
2\omega^{\prime}_{crit}N^{\prime\prime}.$$

On the other hand, from the expansion for $N$  we  have
$$\rho - \rho_{crit} =
\pm \left(\left(- {2\over N^{\prime\prime}}\right)
(N_{crrit} - N)\right)^{1\over 2}  +
\left(N^{\prime\prime\prime} \over
3{N^{\prime\prime}}^2\right)(N_{crit} - N) + ... .$$
Substituting this expression in the expansion for $M_{T}$ we obtain
\ben
M_{T} = {M_{T}}_{crit}  + {\omega}_{crit}(N - N_{crit})
\mp {1\over 3} {\omega^{\prime}}_{crit}
\left(-2\over N^{\prime\prime} \right)^{1\over 2}(N_{crit} - N)^{3\over 2}
+ O(2)
\la{MTEN}
\een
Let's denote by $M_{T}^{up}$ and  $M_{T}^{low}$  correspondingly the
tensor mass on the upper  and lower branch  of the curve $M_{T}(N).$
Then making use of (\ref{MTEN}) one obtains
\ben
M_{T}^{up} - M_{T}^{low} =
-{2\over 3} {\omega^{\prime}}_{crit}
\left(-2\over N^{\prime\prime}\right)^{1\over 2} (N_{crit} - N)^{3\over2}
\een
It's easy to see that for the binding energy we have the same relation
\ben
E_{B}^{up} - E_{B}^{low} =
-{2\over 3} {\omega^{\prime}}_{crit}
\left(-2\over N^{\prime\prime}\right)^{1\over 2} (N_{crit} - N)^{3\over2}
\een
These relations are  scalar-tensor boson star versions of the similar
relations in the fermion stars theory in pure general relativity \cite{HTWW}.
It should be noted that such dependence  $\sim (N_{crit} - N)^{3\over 2}$
is typical for catastrophe theory \cite{AVGZ}.

\bigskip
\bigskip

\noindent{\Large \bf Conclusion}
\bigskip

Scalar-tensor theories of gravity violate the strong equivalence principle.
This results in the appearance of three different possible masses as a
measure of the total energy of the boson star.The stability analysis of the
boson stars requires that the mass and particle number peak, as functions
of the central density, at the same location.In this article we have proved
analytically  that it's the tensor mass which  possess the desirable property.
Therefore, it's the tensor mass which should be taken as the physical mass,
for example in the construction of the binding energy of the star.
While the numerical calculations have been done for boson stars
in pure Brans-Dicke theory  of gravity,  our proof holds for
the most general scalar-tensor theory.Especially, our proof
involves a potential term for the gravitational scalar $\Phi.$ This is
important, because many scalar-tensor models of gravity involve such term.
On the other hand the potential term may play significant role in the early
universe and to influence the boson star formation and stability,although
at present there aren't numerical investigations of boson stars in
scalar-tensor theories with a potential term.

Finally, we believe that the result in this letter has a more general nature.
It's the tensor mass which should be consider as the best candidate
for the physical mass as a measure  of the total energy of space-time in
scalar-tensor theories of gravity.

\bigskip
\bigskip

\noindent{\Large\bf Acknowledgments}

\bigskip
\bigskip
\bigskip
The author is grateful to P. Fiziev for his continuous encouragement
and valuable comments.

The author is also grateful to the anonymous referees  for their
valuable suggestions.

The work on this letter has been partially supported by
the Sofia University Foundation for Scientific Research,
Contracts~No.No.~245/99,~257/99 .

\end{document}